\documentclass[pdftex,twocolumn,epjc3]{svjour3}          

\RequirePackage[T1]{fontenc}

\smartqed  

\usepackage{amssymb}

\RequirePackage{graphicx}
\RequirePackage{mathptmx}      
\RequirePackage{flushend}
\RequirePackage[numbers,sort&compress]{natbib} 
\RequirePackage{amsmath}
\RequirePackage[colorlinks,citecolor=blue,urlcolor=blue,linkcolor=blue]{hyperref}

\journalname{Eur. Phys. J. C}

\begin{document}

\title{Exciting the vacuum: Quantum particle creation during the binary black hole inspiral}


\author{Sohrab Rahvar\thanksref{e1,addr1}
}

\thankstext{e1}{e-mail: rahvar@sharif.edu}

\institute{Physics Department, Sharif University of Technology, Tehran 11365-9161,  Iran \label{addr1}
}

\date{Received: date / Accepted: date}

\maketitle

\begin{abstract}
We investigate particle production in the dynamical curved spacetime of a binary black hole system. Particle production is a well-known feature of quantum field theory in curved spacetime, underlying the Hawking and Unruh effects. Here we extend it to the time-varying gravitational perturbation sourced by an inspiraling binary black hole. Treating a massless scalar field coupled to the binary metric, we compute the particle flux and radiated energy to leading order in the metric perturbation $h_{\mu\nu}$, using both the Bogoliubov transformation method and the S-matrix formalism. The perturbation is modeled with the standard quadrupole formalism, retaining the time-domain quadrupolar ($\ell=2$) contribution that dominates gravitational-wave emission. Our calculation is rigorously valid in the weak-field, large-separation inspiral regime. In this limit, we find a characteristic non-thermal, power-law emission with $dE/dt \propto \mathcal{M}^{10/3}\omega_{gw}^{16/3}$, in contrast to a thermal Hawking spectrum. Crucially, we find that the intensity of the produced particles in this linear regime is weak. This suggests that generating observable multimessenger signatures would require non-linear enhancements from high-amplitude gravitational waves. Extending this analysis to the strong-field merger phase, which necessitates numerical relativity, is left for future work.
\end{abstract}

\section{Introduction}
In flat Minkowski spacetime, global Poincaré symmetry selects a unique vacuum state, leading to a definition of particles that is unambiguously agreed upon by all inertial observers. In contrast, in curved spacetimes—or for non-inertial observers in flat spacetime—the absence of such global symmetry means no unique, preferred vacuum exists \cite{Birrell,Wald:1995yp}. As a result, when field modes naturally defined in one asymptotic region or coordinate system are compared to those in another, the mismatch in vacuum states generally manifests as particle creation.

Two famous examples of particle creation are Hawking radiation from black holes and the Unruh effect experienced by uniformly accelerated observers\cite{Hawking:1974sw,Hawking:1975vcx,Unruh:1976db}. Both cases demonstrate possible connection between gravity, acceleration, and quantum field theory. Beyond these stationary settings, time-dependent gravitational fields can also induce particle production. A classic example is cosmological particle creation in an expanding universe\cite{Parker:1968mv,2024RvMP...96d5005K}. Similarly, dynamical spacetime perturbations, such as those generated by orbiting compact binaries, can act as a time-dependent background for quantum fields, leading to non-thermal particle creation. This effect can be viewed as a quantum analogue of classical gravitational wave emission: just as oscillating masses emit gravitational radiation, the time-varying curvature can excite quantum fields, producing particles. In contrast to this case, it has been shown that for a stationary plane wave no particle creation happens \cite{gibbons1975quantized}.

In this work for a time-varying metric,  we calculate the particle production rate for a massless scalar field propagating on the spacetime of a binary black hole system with component masses $M_1$ and $M_2$. The binary generates a time-periodic metric perturbation $h_{\mu\nu}$. An observer located in the asymptotic Minkowski region far from the binary will detect a flux of particles resulting from the interaction of the quantum field with this dynamical geometry. Our goal is to compute this particle flux to leading order in the metric perturbation, treating $h_{\mu\nu}$ as a classical background field. The problem combines techniques from post-Newtonian expansions and quantum field theory in curved spacetime. The metric perturbation is evaluated using the standard quadrupole formalism for a binary system. The quantum field is then quantized on this time-dependent background, and the Bogoliubov transformation between asymptotic {\it in} and {\it out} vacua yields the particle production rate. We take the time-domain quadrupolar ($\ell=2$) contribution, which corresponds to the leading gravitational-wave emission. 

We reanalyzed this problem from the standard S-matrix formalism \cite{Garriga:1991} where the {\it in} vacuum state scatters from the dynamical perturbation of the spacetime to an {\it out} state as a particle creation. This part of calculation shows the equivalence of Bogoliubov transformation and S-matrix formalism. Finally we calculate the particle production rate when two blackholes inspiral. The burst of particles alongside with the gravitational wave might be a tool for detection of primordial blackholes (PBHs) in the binary form. 

Here we adopt the convention of $G=c=1$ with the signature of $(-+++)$. The organization of the paper is as follows: In Section~\ref{wf}, we derive the classical, time-dependent metric perturbation generated by a binary black hole system at large distances. In Section~\ref{qf}, we consider a scalar field in this background and quantize it, investigating the particle creation from the gravitational waves.  In Section~\ref{Bogoliubov} we derived the Bogoliubov coefficient. Also we perform our calculation based on S-matrix formalism and compared it with the Bogoliubov formalism . We derive the Bogoliubov coefficients in Section~\ref{sec:bogoliubov}. The numerical results for the flux of particles and also observational investigation is given in Section~\ref{num}. The conclusion is given in Section~\ref{conc}. 

\section{Weak-Field Metric Perturbations in the Radiation Zone}
\label{wf}

The direct observation of gravitational waves from compact binary mergers \cite{Abbott:2016blz} stands as a landmark achievement of general relativity, firmly establishing dynamical spacetimes as an observable reality. Motivated by this, we consider a binary black hole system consisting of two masses, $M_1$ and $M_2$, in a quasicircular orbit. The system is characterized by its total mass $M = M_1 + M_2$. We define the symmetric mass ratio $\nu = \frac{M_1 M_2}{M^2}$ and the chirp mass $\mathcal{M} = \nu^{3/5} M$, which govern the inspiral evolution and will be utilized in subsequent calculations. To evaluate the particle production of a scalar field in the wave zone ($r \gg M$), we first require an analytic expression for the classical dynamics of the metric perturbation \cite{Misner:1974qy,Blanchet:2013haa}.

At large distances from the source, the spacetime metric can be decomposed as
\begin{equation}
    g_{\mu\nu} = \eta_{\mu\nu} + H_{\mu\nu} + h_{\mu\nu} \, ,
    \label{eq:metric_decomp}
\end{equation}
where $\eta_{\mu\nu}$ is the flat Minkowski metric. The term $H_{\mu\nu} \sim \mathcal{O}(M/r)$ represents the static, spherically symmetric background field corresponding to the total mass $M$ of the binary. The term $h_{\mu\nu}$ represents the dynamical gravitational wave perturbation generated by the orbital motion of the masses.

Because our analysis takes place in the asymptotic radiation zone, we impose the Transverse-Traceless (TT) gauge on the perturbation $h_{\mu\nu}$, ensuring it is purely spatial, divergenceless, and traceless:
\begin{equation}
    h_{tt}^{TT} = 0, \quad h_{ti}^{TT} = 0, \quad \partial^j h_{ij}^{TT} = 0, \quad \delta^{ij} h_{ij}^{TT} = 0 \, .
\end{equation}
Linearizing the vacuum Einstein equations, $R_{\mu\nu}=0$, for this metric yields the propagation equation for the gravitational wave:
\begin{equation}
    \left( -\partial_t^2 + \nabla^2 \right) h_{ij}^{TT} - \frac{4M}{r} \partial_t^2 h_{ij}^{TT} \approx 0 \, ,
    \label{eq:wave_eq}
\end{equation}
where $\nabla^2$ is the flat-space Laplacian.

The solutions to this equation are naturally governed by the retarded time $t_{ret} = t - r_*$, where $r_*$ is defined as 
    $r_* = r + 2M \ln\left(\frac{r}{2M}-1\right)$  and the generic solution is as $h = F(t-r^\star) + G(t+r^\star)$ \cite{Maggiore:2007ulw}.

At leading order, the gravitational wave strain at a large distance $r$ is determined entirely by the second order time derivative of the transverse-traceless projection of the binary's mass quadrupole moment $I_{ij}$ as the source, evaluated at the retarded time:
\begin{equation}
    h_{ij}^{TT}(t, r) \simeq \frac{2}{r} \ddot{I}_{ij}^{TT}(t - r_*) \, .
    \label{eq:quadrupole}
\end{equation}

In the next section our aim is to calculate the Bogoliubov coefficients, therefore, we require an explicit representation of the gravitational wave.  By evaluating the quadrupole moment for a quasicircular binary, the spatial strain propagating in the radiation zone can be written in terms of two fundamental polarization states:
\begin{equation}
    h_{ij}^{TT}(t, r) = \frac{1}{r} \left[ A_+(\tau) e_{ij}^{+} \cos(\Phi(\tau)) + A_\times(\tau) e_{ij}^{\times} \sin(\Phi(\tau)) \right] \, ,
    \label{eq:time_domain_strain}
\end{equation}
where $e_{ij}^{+,\times}$ represent the constant polarization tensors bases, which depends on the orientation of the binary relative to the observer, and $\tau = t_c - t_{ret}$ is the time remaining until coalescence.

The amplitudes $A_+$ and $A_\times$ in Eq.~\ref{eq:time_domain_strain} are explicitly determined by the binary parameters and the inclination angle $\theta$ of the orbital plane with respect to the line of sight \cite{Maggiore:2007ulw}:
\begin{align}
    A_+(\tau) &= 4 \mathcal{M}^{5/3} \left( \frac{\omega_{gw}(\tau)}{2} \right)^{2/3} \frac{1+\cos^2\theta}{2} \, , \label{eq:amplitude_plus} \\
    A_\times(\tau) &= 4 \mathcal{M}^{5/3} \left( \frac{\omega_{gw}(\tau)}{2} \right)^{2/3} \cos\theta .  \label{eq:amplitude_cross}
\end{align}

We note that $\omega_{gw}(\tau)$ is the instantaneous gravitational wave frequency. As the binary system loses energy through the emission of gravitational radiation, the orbit shrinks and the orbital frequency increases. The evolution of the frequency is dictated entirely by the chirp mass $\mathcal{M}$. Integrating the leading-order PN radiation yields the frequency as a function of the remaining time as
\begin{equation}  
 \omega_{gw}(\tau) = 2\left( \frac{5}{256 \, \mathcal{M}^{5/3} \tau} \right)^{3/8}
\label{omegat}
\end{equation}
 and integrating the frequency over time provides the explicit time-domain gravitational wave phase: $$\Phi(\tau) = \phi_c - 2 \left( \frac{\tau}{5 \mathcal{M}} \right)^{5/8}$$  where $\phi_c$ is the phase of coalescence. 

We can substitute $\omega_{gw}(\tau)$ into the general amplitude expressions (Eqs.~\ref{eq:amplitude_plus} and \ref{eq:amplitude_cross}) and write the proportional time-dependent growth of the generic overall amplitude up to the time of coalescence ($t \to t_c$, or $\tau \to 0$):
\begin{equation}
    A_{+,\times}(\tau) \propto \mathcal{M}^{5/3} \left[ \omega_{gw}(\tau) \right]^{2/3} = \mathcal{M}^{5/3} \left( \frac{5}{256 \, \mathcal{M}^{5/3} \tau} \right)^{1/4} \, .
    \label{eq:time_amplitude}
\end{equation}

By substituting the time-dependent phase $\Phi(\tau)$ and the amplitudes from Eq.~\ref{eq:time_amplitude} directly into Eq.~\ref{eq:time_domain_strain}, we obtain a fully analytical classical solution for the gravitational waves at large distances in terms of time.

\section{Quantum Field Theory in Dynamical Spacetimes}
\label{qf}

In this section, we consider a massless scalar field $\phi$ that satisfies $\Box_g \phi = 0$. While the full spacetime includes the curvature generated by the total mass $M$, we focus our evaluation of the scattering amplitude in the asymptotic region. There, the metric can be approximated as $g_{\mu\nu} = \eta_{\mu\nu} + H^{\mu\nu} + h^{TT}_{\mu\nu}$, expanding around a flat Minkowski background $\eta_{\mu\nu}$. Consequently, the d'Alembertian splits as $\Box_g = \Box^{(0)} + \delta\Box$, where $\Box^{(0)}$ is the wave operator in Schwarzschild metric at asymptotically distance from the central binary black hole (i.e. $\eta_{\mu\nu} + H^{\mu\nu}$). $\delta\Box$ contains terms linear in the transverse-traceless gravitational wave perturbation $h_{\mu\nu}^{TT}$.

Now we expand the field in modes that are purely incoming from past null infinity ($\mathcal{I}^-$) and purely outgoing to future null infinity ($\mathcal{I}^+$):
\begin{align}
\phi^{\text{in}}(x) &= \sum_{\ell,m} \int_0^\infty d\omega \left( a_{\omega\ell m}^{\text{in}} u_{\omega\ell m}^{\text{in}}(x) + \text{h.c.} \right), \\
\phi^{\text{out}}(x) &= \sum_{\ell,m} \int_0^\infty d\omega \left( a_{\omega\ell m}^{\text{out}} u_{\omega\ell m}^{\text{out}}(x) + \text{h.c.} \right).
\end{align}
The asymptotic solutions for the differential equation (i.e., $\Box^{(0)}\phi =0$) are \footnote{See Chapter 8 of \cite{Birrell}}
\begin{align}
u_{\omega\ell m}^{\text{in}}(x) &\sim \frac{1}{\sqrt{4\pi\omega}} \frac{e^{-i\omega v}}{r} Y_{\ell m}(\theta,\phi), \quad v=t+r_*, \\
u_{\omega\ell m}^{\text{out}}(x) &\sim \frac{1}{\sqrt{4\pi\omega}} \frac{e^{-i\omega u}}{r} Y_{\ell m}(\theta,\phi), \quad u=t-r_*,
\end{align}
with the tortoise coordinate $r_* = r + 2M\ln(r/(2M)-1)$. In order to calculate the particle production, we use these "in" and "out" solutions as the basis for our calculations. 

\subsection{Vacuum Ambiguity and Particle Production}

To rigorously quantify the particle production induced by the binary black hole inspiral, we must address the concept of vacuum ambiguity in curved and dynamical spacetimes. In standard flat Minkowski spacetime, global Poincaré symmetry allows for a unique, universal definition of the vacuum state and particle number; technically speaking, the spacetime possesses a global timelike Killing vector. However, in our system, the presence of both the static background curvature (due to the total mass $M$) and the time-dependent gravitational wave perturbation $h_{ij}^{TT}(t,r)$ breaks this global time-translation symmetry. 

\begin{figure}[t]
\centering
\includegraphics[width=\columnwidth]{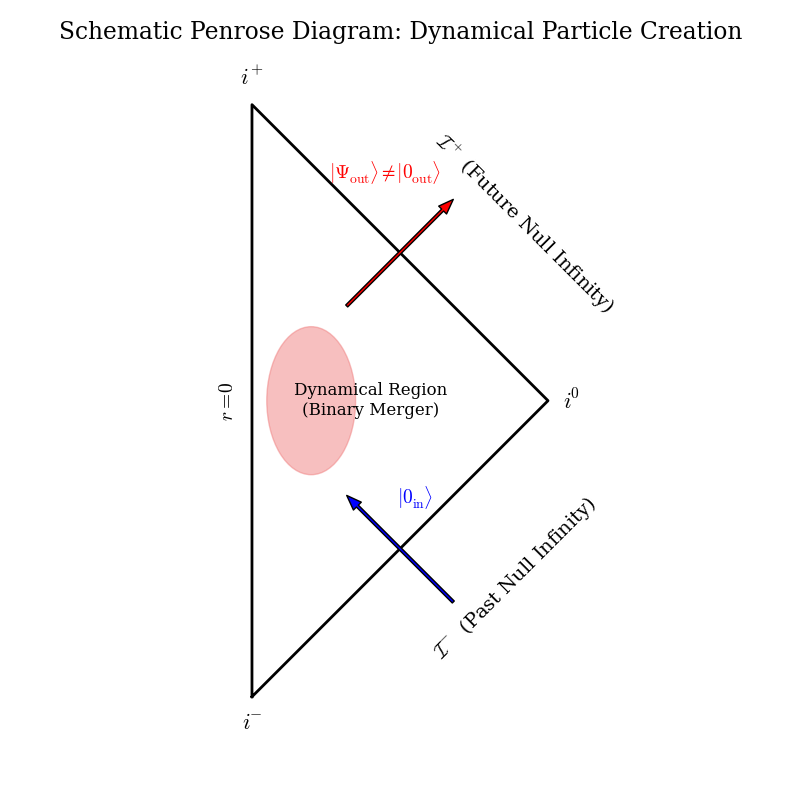}
\caption{Schematic Penrose diagram illustrating the "in" and "out" states.}
\label{figure1}
\end{figure}

Because the spacetime is dynamical, a state devoid of particles at early times will not generally be perceived as a vacuum at late times. We therefore define two asymptotic regions where the spacetime settles, allowing us to define unambiguous particle states. The "in" region at past null infinity ($\mathcal{I}^-$) corresponds to early times ($t \to -\infty$) when the binary components are widely separated, the orbital velocity is low, and the metric is approximately stationary. The second state is the "out" region at future null infinity ($\mathcal{I}^+$), corresponding to late times after the merger, where observers are equipped with particle detectors (see Figure \ref{figure1}). These two different states are analogous to S-matrix theory, where one considers distinct asymptotic "in" and "out" states. We can now derive the probability of a specific scattering process where the vacuum state transitions into a multi-particle state.

Let $\hat{\phi}(x)$ be the quantum scalar field operator. We can expand this field in terms of a complete set of orthonormal mode functions $u_J^{\text{in}}(x)$ adopted to the "in" region at $t \to -\infty$, where $J$ denotes the relevant quantum numbers:
\begin{equation}
\hat{\phi}(x) = \sum_J \left( \hat{a}_J^{\text{in}} u_J^{\text{in}}(x) + \hat{a}_J^{\text{in}\dagger} u_J^{\text{in}*}(x) \right).
\label{eq:in_expansion}
\end{equation}
The operators $\hat{a}_J^{\text{in}}$ and $\hat{a}_J^{\text{in}\dagger}$ are the annihilation and creation operators for the "in" states. The "in" vacuum represents the initial absence of quantum scalar particles and is defined by:
\begin{equation}
\hat{a}_J^{\text{in}} |0_{\text{in}}\rangle = 0 \quad \text{for all } J.
\end{equation}
In other words, $|0_{\text{in}}\rangle$ is empty of quantum scalar particles within the classical gravitational background generated by the binary black holes with masses $M_1$ and $M_2$ at $t \to -\infty$.

Similarly, we can expand the same field $\hat{\phi}(x)$ using a different set of mode functions $u_K^{\text{out}}(x)$ adopted to the "out" region:
\begin{equation}
\hat{\phi}(x) = \sum_K \left( \hat{a}_K^{\text{out}} u_K^{\text{out}}(x) + \hat{a}_K^{\text{out}\dagger} u_K^{\text{out}*}(x) \right).
\label{eq:out_expansion}
\end{equation}
Observers at future infinity define their vacuum $|0_{\text{out}}\rangle$ and their particle number operator using these "out" modes:
\begin{equation}
\hat{N}_K^{\text{out}} = \hat{a}_K^{\text{out}\dagger} \hat{a}_K^{\text{out}}.
\end{equation}

Because both sets of mode functions $\{u_J^{\text{in}}\}$ and $\{u_K^{\text{out}}\}$ form complete bases for solutions to the wave equation $\Box_g \phi = 0$, the "out" operators can be expressed in terms of the "in" operators via Bogoliubov transformations:
\begin{equation}
\hat{a}_K^{\text{out}} = \sum_J \left( \alpha_{KJ}^* \hat{a}_J^{\text{in}} - \beta_{KJ}^* \hat{a}_J^{\text{in}\dagger} \right),
\label{eq:bogoliubov_operators}
\end{equation}
where $\alpha_{KJ}$ and $\beta_{KJ}$ are the Bogoliubov coefficients. They are defined via the inner product between the complete bases $\{u_J^{\text{in}}\}$ and $\{u_K^{\text{out}}\}$ as $\alpha_{KJ} = (u_K^{\text{out}}, u_J^{\text{in}})$ and $\beta_{KJ} = - (u_K^{\text{out}}, u_J^{\text{in}*})$, utilizing the inner product property $(u^*,v^*) = -(u,v)^*$. The covariant Klein-Gordon inner product of two complex scalar fields $f$ and $g$ evaluated on a spacelike Cauchy surface $\Sigma$ is defined as:
\begin{equation} 
(f, g)_\Sigma = i \int_{\Sigma} d^3x \sqrt{\gamma} \, n^\mu \left( f^* \nabla_\mu g - g \nabla_\mu f^* \right),
\label{inn}
\end{equation}
where $\gamma$ is the determinant of the induced metric on $\Sigma$, and $n^\mu$ is the future-directed unit normal vector to the hypersurface. 

Using the definition of the Bogoliubov coefficients, we can also write the complete basis of the "out" region in terms of the "in" region as follows:
\begin{equation}
u_K^{\text{out}}(x) = \sum_J \left( \alpha_{KJ} u_J^{\text{in}}(x) + \beta_{KJ} u_J^{\text{in}*}(x) \right).
\label{eq:bogoliubov_modes}
\end{equation}

Equation (\ref{eq:bogoliubov_operators}) represents the core mathematical mechanism of particle creation. We start in the state $|0_{\text{in}}\rangle$ prior to the intense gravitational wave emission. To determine what the late-time observer detects, we must evaluate the expectation value of their particle counter, $\hat{N}_K^{\text{out}}$, in the state the universe actually started in ($|0_{\text{in}}\rangle$). 

Substituting Eq.~\eqref{eq:bogoliubov_operators} into the expectation value yields:
\begin{align}
\langle 0_{\text{in}}| \hat{N}_K^{\text{out}} |0_{\text{in}}\rangle &= \langle 0_{\text{in}}| \hat{a}_K^{\text{out}\dagger} \hat{a}_K^{\text{out}} |0_{\text{in}}\rangle \nonumber \\
&= \langle 0_{\text{in}}| \left[ \sum_{J'} \left( \alpha_{KJ'} \hat{a}_{J'}^{\text{in}\dagger} - \beta_{KJ'} \hat{a}_{J'}^{\text{in}} \right) \right] \nonumber \\ 
&\times \left[ \sum_J \left( \alpha_{KJ}^* \hat{a}_J^{\text{in}} - \beta_{KJ}^* \hat{a}_J^{\text{in}\dagger} \right) \right] |0_{\text{in}}\rangle.
\end{align}
Because $\hat{a}_J^{\text{in}} |0_{\text{in}}\rangle = 0$, any term where an annihilation operator acts to the right vanishes. Similarly, $\langle 0_{\text{in}}| \hat{a}_{J'}^{\text{in}\dagger} = 0$, eliminating terms where a creation operator acts to the left. The only surviving term comes from the product of the two $\beta$ terms. Utilizing the commutation relation $[\hat{a}_J^{\text{in}}, \hat{a}_{J'}^{\text{in}\dagger}] = \delta_{JJ'}$, we obtain:
\begin{equation}
\langle 0_{\text{in}}| \hat{N}_K^{\text{out}} |0_{\text{in}}\rangle = \sum_J |\beta_{KJ}|^2.
\label{eq:dynamical_production}
\end{equation}

If the spacetime were perfectly static or stationary, it would possess a global timelike Killing vector field. This symmetry allows for a universal, unambiguous definition of time across the entire spacetime, meaning the wave equation solutions can be globally and cleanly separated into strictly positive frequency modes (proportional to $e^{-i\omega t}$ with $\omega>0$) and negative frequency modes (proportional to $e^{+i\omega t}$). Because the Klein-Gordon inner product between any positive frequency mode and any negative frequency mode is exactly zero, positive and negative frequencies do not mix. Consequently, we would have $\beta_{KJ} = 0$ which is non-zero in our case. 

The expectation value in Eq.~\eqref{eq:dynamical_production} represents a direct conversion of the classical gravitational wave energy into discrete quantum excitations observed at infinity. Transitioning the summation into a continuous angular frequency notation, the total particle production for a given "out" mode is evaluated by integrating the Bogoliubov coefficient over the continuous "in" spectrum:
\begin{equation}
\langle 0_{\text{in}}| N_{\text{out}}(\omega, J) |0_{\text{in}}\rangle = \int d\omega' \sum_{J'} |\beta_{\omega, J, \omega', J'}|^2.
\label{eq:dynamical_production_continuous}
\end{equation}
We emphasize that the following calculation evaluates the leading-order contribution in the metric perturbation $h_{\mu\nu}$. In this perturbative framework, the relevant Bogoliubov coefficient describes the overlap between the exact physical field (which evolves from the initial vacuum) and the unperturbed background mode $u_{out}$. Projecting onto these background modes is the standard technique for computing the first-order transition amplitude. Sub-leading contributions from the near-zone curvature and higher multipoles are neglected as they are of higher order in the weak-field post-Newtonian expansion.

\section{Derivation of the Bogoliubov Coefficient as a Volume Integral}
\label{Bogoliubov}

To calculate the Bogoliubov coefficient, we must project the quantum states defined in the asymptotic past (the ``in'' region) onto those in the asymptotic future (the ``out'' region). Let us use $\beta^*$ instead of $\beta$ for simplicity in calculation, where $\beta^* = - \left( u_{\text{out}}^*, u_{\text{in}} \right)_{\Sigma_{\text{out}}} $ on a Cauchy future surface $\Sigma_{\text{out}}$ at $t \rightarrow \infty$. Here, $u_{\text{in}}$ represents the exact physical field that evolved from the past vacuum, and $u_{\text{out}}$ is the unperturbed reference mode in the future. Also, note that for simplicity in notation, we have dropped the indices of $J$ and $K$ for $\beta$ in this part. 

Evaluating this directly at $t \rightarrow \infty$ is analytically challenging because the physical field $\phi_{\text{in}}$ has been distorted by evolving through the dynamic spacetime of the binary merger. To circumvent this, we convert the surface integral into a volume integral. Based on the definition of the background Klein-Gordon inner product, we define the current:
\begin{equation}
J^\mu = i \left( u_{\text{out}} \nabla_{(0)}^\mu u_{\text{in}} - u_{\text{in}} \nabla_{(0)}^\mu u_{\text{out}} \right).
\end{equation}
Note that we use $u_{\text{out}}$ rather than its complex conjugate because the inner product $(u_{\text{out}}^*, \phi_{\text{in}})$ involves taking the conjugate of the first argument, giving $(u_{\text{out}}^*)^* = u_{\text{out}}$. 

We recognize that this is precisely the standard Noether current for the pair of fields $(u_{\text{out}}^*, u_{\text{in}})$. Unlike the exact case where both fields satisfy the same equation of motion, this current is not conserved here because $u_{\text{in}}$ satisfies the full perturbed equation. Indeed, its divergence yields the interaction source term:
\begin{equation}
\label{J}
\nabla^{(0)}_\mu J^\mu = i\left( u_{\text{out}} \Box^{(0)} u_{\text{in}} - u_{\text{in}} \Box^{(0)} u_{\text{out}} \right) = -i\, u_{\text{out}} \,\delta\Box\, u_{\text{in}} \sim \mathcal{O}(h),
\end{equation}
assuming that $u_{in}$ satisfies the full equation of motion, $\Box_g u_{\text{in}} = (\Box^{(0)} + \delta\Box)u_{\text{in}} = 0$.

The integral of this current over any spatial hypersurface $\Sigma$ gives the inner product $-\left( u_{\text{out}}^*, u_{\text{in}} \right)_{\Sigma}$. Let us consider the past boundary $\Sigma_{\text{in}}$ (as $t \rightarrow -\infty$). In this remote past region, the dynamical perturbation vanishes: $h_{\mu\nu} \to 0$. Consequently, the exact field $u_{\text{in}}$ reduces identically to a positive-frequency solution of the background equation. The mode $u_{\text{out}}$ used in our projection is, by construction, the unperturbed background reference mode, which is also purely positive-frequency. Since positive-frequency modes are strictly orthogonal under the inner product $(f^*, g)$, the surface integral on $\Sigma_{\text{in}}$ evaluates to zero. 

We apply Gauss's theorem to a spacetime volume $V$ bounded by $\Sigma_{\text{in}}$ and $\Sigma_{\text{out}}$. Because the Bogoliubov coefficients are defined via projection onto the unperturbed background modes in this leading-order perturbative framework, we apply Gauss's theorem using the background Schwarzschild metric $g^{(0)}_{\mu\nu}$:
\begin{equation}
\int_V d^4x \sqrt{-g^{(0)}} \nabla^{(0)}_\mu J^\mu = \int_{\Sigma_{\text{out}}} J^\mu d\Sigma_\mu - \int_{\Sigma_{\text{in}}} J^\mu d\Sigma_\mu,
\end{equation}
where $d\Sigma_\mu$ is the surface element induced by the background metric. Substituting our boundary evaluations, the right-hand side simplifies to $-\beta^*$. This leaves us with:
\begin{equation}
\beta^* = -\int_V d^4x \sqrt{-g^{(0)}} \nabla^{(0)}_\mu J^\mu.
\end{equation}

Working to first order in the metric perturbation (the Born approximation) and substituting equation (\ref{J}), results in 
\begin{equation}
\beta^* = i \int_V d^4x \sqrt{-g^{(0)}} \, u_{\text{out}} \, \delta\Box \, u_{\text{in}},
\label{b1}
\end{equation}
where it is understood that $u_{\text{in}}$ on the right-hand side is evaluated at zeroth order.

Because the observable particle number is proportional to $|\beta|^2$, this integral fully dictates the particle production. Before calculating this integral explicitly, it is instructive to investigate the particle production from the complementary perspective of S-matrix theory.

\subsection{Particle Creation via the S-Matrix Formalism}

In this section, we demonstrate the equivalence between the Bogoliubov transformation and the first-order S-matrix description of particle creation. The essential physical idea is that the time-dependent gravitational perturbation generated by the binary system acts as an external classical source, exciting the quantum vacuum and producing pairs of particles.

We formulate the theory on the background spacetime described by the metric $g^{(0)}_{\mu\nu}$, while the full metric is written as
\begin{equation}
g_{\mu\nu}(x)=g^{(0)}_{\mu\nu}(x)+h_{\mu\nu}(x),
\end{equation}
where $h_{\mu\nu}$ represents the dynamical perturbation sourced by the binary black hole system.

The action for a massless scalar field $\hat{\phi}$ in this spacetime is
\begin{equation}
{\cal S} = -\frac12 \int d^4x \sqrt{-g}\, g^{\mu\nu} \nabla_\mu \hat{\phi} \nabla_\nu \hat{\phi}.
\end{equation}

Expanding the metric to first order in the perturbation gives
\begin{align}
\sqrt{-g} &\simeq \sqrt{-g^{(0)}} \left( 1+\frac12 g^{(0)\alpha\beta}h_{\alpha\beta} \right), \\
g^{\mu\nu} &\simeq g^{(0)\mu\nu}-h^{\mu\nu},
\end{align}
where indices on $h^{\mu\nu}$ are raised and lowered with the background metric. The action separates into a free part and an interaction part,
\begin{equation}
{\cal S}={\cal S}_{\rm free}+{\cal S}_{\rm int}.
\end{equation}

To linear order in $h_{\mu\nu}$, the interaction action becomes
\begin{equation}
{\cal S}_{\rm int} = -\frac12 \int d^4x \sqrt{-g^{(0)}} \left( \frac12 h\, g^{(0)\mu\nu} - h^{\mu\nu} \right) \nabla^{(0)}_\mu \hat{\phi} \nabla^{(0)}_\nu \hat{\phi},
\end{equation}
where $h=g^{(0)\alpha\beta}h_{\alpha\beta}$. In the transverse-traceless gauge the trace vanishes, $h=0$.

Now we define the interaction Hamiltonian as
\begin{equation}
\hat{H}_{\rm int}(t)
= - \int d^3x\,\hat{\mathcal L}_{\rm int},
\end{equation}
and the first-order S-matrix operator is
\begin{equation}
\hat{S} \simeq \hat{I} - i \int_{-\infty}^{+\infty} dt\,\hat{H}_{\rm int}(t)
= \hat{I} + i \int d^4x\,\hat{\mathcal L}_{\rm int}.
\end{equation}

The physical process of interest is the transition from the initial vacuum state to a final two-particle state, (i.e. $|0_{\rm in}\rangle \rightarrow |1_{{\rm out},j},1_{{\rm out},k}\rangle$.
Then the corresponding transition amplitude is 
\begin{equation}
\mathcal A_{jk} = \langle 1_{{\rm out},j}, 1_{{\rm out},k} | \hat S | 0_{\rm in} \rangle .
\end{equation}

To evaluate this amplitude, we expand the quantum field operator in asymptotic modes,
\begin{equation}
\hat{\phi}(x)
= \sum_n \left( \hat a_n u_n(x) + \hat a_n^\dagger u_n^*(x) \right).
\end{equation}
Because the initial state is the vacuum, only the terms containing two creation operators,
$\hat a^\dagger \hat a^\dagger$, contribute to the matrix element. Consequently, the pair-production amplitude is obtained from the contraction of two complex-conjugated mode functions:
\begin{equation}
\mathcal A_{jk} \propto i \int d^4x \sqrt{-g^{(0)}} \, h^{\mu\nu} \nabla^{(0)}_\mu u_{{\rm out},j}^* \nabla^{(0)}_\nu u_{{\rm out},k}^* .
\label{eq:Smatrix_pair}
\end{equation}

This expression describes the creation of two scalar particles from the vacuum by the external time-dependent gravitational field. The classical perturbation $h_{\mu\nu}$ injects energy and momentum into the quantum field, analogous to particle production by a classical external source in ordinary quantum field theory.

We now connect this result with the Bogoliubov formalism. The Bogoliubov coefficient is defined through the overlap of positive- and negative-frequency solutions:
\begin{equation}
\beta_{jk}^* = - \left( u_{{\rm out},j}^*, u_{{\rm in},k} \right).
\end{equation}

Equation~(\ref{eq:Smatrix_pair}) and Eq.~(\ref{b1}) represent two equivalent descriptions of the same physical phenomenon. In the S-matrix picture, the external gravitational perturbation creates particle pairs directly from the vacuum. In the Bogoliubov description, the time-dependent geometry mixes positive- and negative-frequency modes, leading to a non-vanishing coefficient $\beta_{jk}$. The coefficient multiplying the $\hat a^\dagger \hat a^\dagger$ term in the S-matrix expansion is precisely the Bogoliubov coefficient, establishing the equivalence between the two approaches. In the next section we calculate the Bogoliubov coefficient for this system and one can show that 
two approach are mathematically identical. 

\begin{figure}[t]
\centering
\includegraphics[width=\columnwidth]{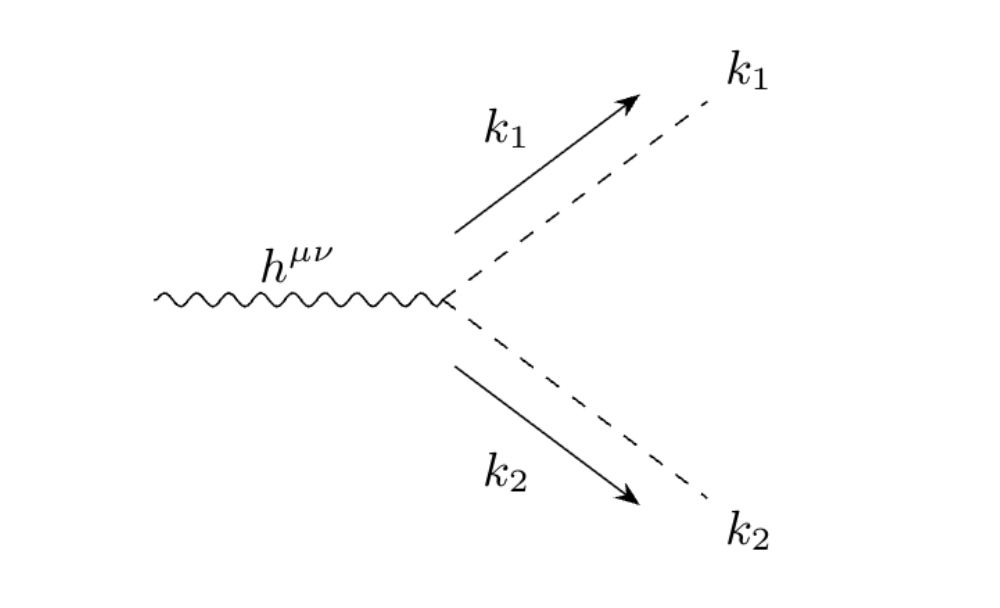}
\caption{Feynman diagram representing particle pair creation. The external wavy line is not a quantum graviton, but the classical gravitational perturbation $h_{\mu\nu}$ carrying four-momentum $q$. It acts as a source for the scalar field (solid lines), creating a pair of particles with four-momenta $k_1$ and $k_2$ such that $q = k_1 + k_2$.}
\label{fig2}
\end{figure}

The corresponding process of particle production is illustrated schematically in Fig.~\ref{fig2}, where the classical gravitational perturbation transfers four-momentum $q^\mu$ into the scalar field and produces a particle pair with momenta $k_1^\mu$ and $k_2^\mu$ satisfying
\begin{equation}
q^\mu = k_1^\mu + k_2^\mu .
\end{equation}

\section{First-Order Bogoliubov Coefficients from the Chirping Binary}
\label{sec:bogoliubov}

To calculate the particle production, we evaluate the Bogoliubov coefficient $\beta_{ij}$, which quantifies the mixing of positive and negative frequency modes due to the time-dependent gravitational background. 

As derived from equation (\ref{b1}), the volume integral for the $\beta$ coefficient requires the complex conjugate of both the ``in'' and ``out'' modes to capture pair creation from the vacuum:
$$ \beta_{\omega\ell m,\,\omega'\ell'm'} = i \int d^4x \sqrt{-g^{(0)}} \; u_{\omega\ell m}^{\text{out}*}(x) \; \delta\Box \; u_{\omega'\ell'm'}^{\text{in}*}(x). $$

Here for simplicity we adopt $\beta$ instead of calculating $\beta^\star$. We established in Section \ref{wf} that the gravitational wave is in the Transverse-Traceless (TT) gauge ($h=0$, $h_{0\mu}=0$), meaning the perturbation to the d'Alembertian simplifies entirely to spatial derivatives:
$$ \delta\Box = h^{ij}_{TT}(t,r) \partial_i \partial_j + \mathcal{O}(h^2). $$

Substituting this into the integral gives:
$$ \beta_{\omega\ell m,\,\omega'\ell'm'} = i \int d^4x \sqrt{-g^{(0)}} \; u_{\omega\ell m}^{\text{out}*} \left( h^{ij}_{TT}(t,r) \partial_i \partial_j u_{\omega'\ell'm'}^{\text{in}*} \right). $$

To reveal the underlying symmetry between the frequencies $\omega$ and $\omega'$, we perform integration by parts on the spatial derivatives. Assuming the boundary term at spatial infinity vanishes and utilizing the transversality condition of the TT gauge ($\nabla_i h^{ij}_{TT} = 0$), the derivative operator is distributed symmetrically across both modes:
$$ \beta_{\omega\ell m,\,\omega'\ell'm'} = -i \int d^4x \sqrt{-g^{(0)}} \; h^{ij}_{TT}(t,r) \left( \partial_i u_{\omega\ell m}^{\text{out}*} \right) \left( \partial_j u_{\omega'\ell'm'}^{\text{in}*} \right), $$
which is equivalent to the S-matrix formalism in equation (\ref{eq:Smatrix_pair}). 

Because positive-frequency modes have a time dependence of $e^{-i\omega t}$, their complex conjugates carry a factor of $e^{+i\omega t}$. By separating the time and spatial domains, the integral becomes:
\begin{eqnarray}
 \beta_{\omega\ell m,\,\omega'\ell'm'} &=& -i \int_{-\infty}^{\infty} dt \int d^3x \sqrt{-g^{(0)}} \; h^{ij}_{TT}(t,r)\\  \nonumber
&&  \partial_i \left( \frac{e^{i\omega u}}{r\sqrt{4\pi\omega}} Y_{\ell m}^* \right) \partial_j \left( \frac{e^{+i\omega' v}}{r\sqrt{4\pi\omega'}} Y_{\ell' m'}^* \right) ,
\end{eqnarray}
where $u = t - r_*$ and $v = t + r_*$. 

In a system with a strictly periodic orbit, $h_{ij}^{TT}(t,r)$ oscillates with a constant frequency $\Omega$, the time integral would yield a Dirac delta function $\delta(\omega+\omega'-\Omega)$, indicating discrete resonant particle creation. However, a binary black hole system is not strictly periodic; it emits energy and chirps. The time-domain strain can be written in complex exponential form as:
$$ h_{ij}^{TT}(t,r) \propto A(t) e^{-i\Phi(t)} e_{ij} + \text{c.c.}, $$
where $A(t)$ is the slowly varying amplitude of gravitational wave and $\Phi(t)$ is the accumulating phase (i.e., $\Phi(t) = \int \omega(t) dt$). Isolating the time-dependent portion of the integral, we have:
$$ I_t = \int_{-\infty}^{\infty} dt \; A(t) \exp\Big[i \big( (\omega + \omega')t - \Phi(t) \big) \Big]. $$

Because the phase factor oscillates rapidly, the integral evaluates to nearly zero everywhere except where the phase is stationary. We define the total phase function $\Psi(t) = (\omega + \omega')t - \Phi(t)$. The stationary phase approximation dictates that the dominant contribution occurs at the time $t_*$ where the derivative of the phase vanishes:
$$ \frac{d\Psi}{dt} = 0 \quad \implies \quad \omega + \omega' = \frac{d\Phi(t_*)}{dt} \equiv \omega_{gw}(t_*). $$
This establishes a critical resonance condition, which is a direct manifestation of energy conservation: the creation of a particle pair with energies $\omega$ and $\omega'$ predominantly occurs at the specific retarded time $t_*$ when the instantaneous  gravitational wave frequency $\omega_{gw}(t_*)$ exactly matches the sum of the created particles' energies.

Now, we can evaluate the integral at $t_*$, by Taylor expansion of $\Psi(t)$ around $t_*$ as $\Psi(t) = \Psi(t_*)+1/2\ddot{\Psi}(t_*)(t-t_*)^2$, which yields:
$$ I_t \approx A(t_*) \sqrt{\frac{2\pi}{|\ddot{\Phi}(t_*)|}} \exp\Big[i \big( (\omega + \omega')t_* - \Phi(t_*) \pm \frac{\pi}{4} \big) \Big]. $$
This term is directly proportional to the Fourier transform of the chirping waveform, $\tilde{h}(\omega_{gw})$, evaluated at the resonance frequency $\omega_{gw} = \omega + \omega'$. Substituting $\ddot\Phi = \dot\omega_{gw}$ and utilizing Post-Newtonian theory, the Fourier amplitude scales with $A(t_\star) \propto \omega_{gw}^{2/3}$ and $\dot\omega_{gw} \propto \omega_{gw}^{11/3}$ \cite{Maggiore:2007ulw}. Therefore, the overall amplitude behaves as $I_t \propto \mathcal{M}^{5/6} \omega_{gw}^{-7/6}$.

With the time integral resolved, the remaining spatial integration acts on the radial mode functions and the angular spherical harmonics. Applying the large-$r$ approximation to the symmetric spatial integral, the derivatives acting on the outgoing and ingoing mode functions pull down their respective radial momenta ($\partial_i \approx -i\omega \hat{r}_i$ and $\partial_j \approx -i\omega' \hat{r}_j$). This produces a factor of $-\omega\omega'$. When combined with the canonical scalar field normalizations extracted from the denominator ($\frac{1}{\sqrt{\omega}}$ and $\frac{1}{\sqrt{\omega'}}$), we obtain the symmetric prefactor $\frac{\omega\omega'}{\sqrt{\omega\omega'}} = \sqrt{\omega\omega'}$. 

Bundling the remaining terms into a reduced spatial overlap integral, and ensuring the correct standard normalization, the purely spatial integration simplifies to:
$$ \mathcal{R}_{\ell m}^{\ell'm'}(\omega,\omega') = -\sqrt{\omega\omega'} \int_0^\infty dr \, r \; \mathcal{I}_{\ell m}^{\ell'm'}(r) \; e^{i(\omega \pm \omega')r}, $$ noting that in the radiation-zone, $r^\star \simeq r$. The angular matrix elements $\mathcal{I}_{\ell m}^{\ell'm'}(r)$ also defines as:
$$ \mathcal{I}_{\ell m}^{\ell'm'}(r) = \frac{1}{4\pi} \int d\Omega \; Y_{\ell m}^*(\Omega) \left[ e^{ij}(\Omega) \hat{r}_i \hat{r}_j \right] Y_{\ell' m'}^*(\Omega). $$

Combining the time evaluation and the spatial integrals,  the Bogoliubov coefficient simplifies to:
\begin{equation}
\label{46}
 \beta_{\omega\ell m,\,\omega' \ell' m'} \simeq \mathcal{N} \sqrt{\omega\omega'} \, \mathcal{M}^{5/6} (\omega + \omega')^{-7/6} \; \mathcal{I}_{\ell m}^{\ell' m'} \; \kappa_{\ell}(\omega) \, , 
 \end{equation}
where $\mathcal{N}$ is a normalization constant capturing the angular inclination factors of the binary. The term $\mathcal{M}^{5/6} (\omega + \omega')^{-7/6}$ originates from the Fourier amplitude $\tilde{h}(\omega_{gw})$ evaluated at the resonance condition and $\kappa_{\ell}(\omega)$ is the greybody factor \cite{1983mtbh.book.....C}, which square of it represents the probability amplitude that a mode of frequency $\omega$ successfully transmits through the effective gravitational potential barrier of the central mass $M$ to reach future null infinity $\mathcal{I}^+$.

\section{Particle Number and Energy Flux at Infinity}
\label{num}

From equation (\ref{46}), the probability of particle creation can be written in a simplified form as 
$$ |\beta_{\omega\ell m,\,\omega' \ell' m'}|^2 = C^2 \mathcal{M}^{5/3} \omega \omega' \omega_{gw}^{-7/3}, $$
where we have applied the resonance condition $\omega + \omega' = \omega_{gw}$, and $\omega_{gw}$ at any given instantaneous time is treated as almost constant. Now, we use the definition of the particle number density:
$$ \frac{d^2N}{d\omega d\omega'} = C^2 \mathcal{M}^{5/3} \omega \omega' \omega_{gw}^{-7/3}. $$

Integrating this equation over the intermediate mode frequencies $\omega' \in [0, \omega_{gw} - \omega]$ yields the differential particle number spectrum:
$$ \frac{dN}{d\omega} = \int_0^{\omega_{gw}-\omega} \frac{d^2N}{d\omega d\omega'} d\omega' = \frac{C^2}{2} \mathcal{M}^{5/3} \omega_{gw}^{-7/3} \omega (\omega_{gw} - \omega)^2. $$

To obtain the energy spectrum of the generated particles, we multiply the particle number by the energy of each particle, $\omega$ (setting $\hbar = 1$). Absorbing the factor of $1/2$ into a redefined constant $C^2$, the differential energy spectrum evaluates to:
\begin{equation}
\label{energy_spectrum}
\frac{dE}{d\omega} = \omega \frac{dN}{d\omega} = C^2 \mathcal{M}^{5/3} \omega_{gw}^{-7/3} \omega^2 (\omega_{gw} - \omega)^2.
\end{equation}
Since the frequency of gravitational wave is changing by time, then the rate of energy emission for a frequency would be differentiating this equation with respect to the time

$$ \frac{d^2E}{dt d\omega} = \frac{1}{3} C^2 \mathcal{M}^{5/3} \omega^2 \omega_{gw}^{-10/3} (\omega_{gw} - \omega)(7\omega - \omega_{gw}) \dot{\omega}_{gw} $$ 
substituting $\dot\omega$, we finally obtain this expression in terms of the $\omega$ and $\omega_{gw}$. 
$$ \frac{d^2E}{dt d\omega} \propto \mathcal{M}^{10/3} \omega_{gw}^{1/3} \omega^2 (\omega_{gw} - \omega)(7\omega - \omega_{gw}) $$
and integrating over $\omega$ we get the following experssion for the power of emission of particles 

$$ \frac{dE}{dt} \propto \mathcal{M}^{10/3} \omega_{gw}^{16/3} $$
Now, we use $\hbar$, $G$ and $c$ and apply the dimensional analysis. Here, the expression for the power of the emission as 
\begin{equation}
\label{power}
\frac{dE}{dt} = P = N\hbar \left( \frac{G \mathcal{M}}{c^3} \right)^{10/3} \omega_{gw}^{16/3},
\end{equation}
where $N$ is the normalization factor and that is dimensionless. Assuming $N$ in the order of unity, we use the numerical values for the constant 
the final result is 
$$ \frac{dE}{dt} \sim  10^{-30} \left( \frac{\mathcal{M}}{M_\odot} \right)^{10/3} \left( \frac{f_{gw}}{100 \text{ Hz}} \right)^{16/3} \left[ \frac{\text{erg}}{\text{s}} \right]. $$
Since the binary sweeps through frequency as it inspirals, differentiating the total power (\ref{power}) with respect to the frequency yields the differential power spectrum.
\begin{equation}
 \frac{dP}{df_{gw}} \sim  5\times 10^{-32} \left( \frac{\mathcal{M}}{M_\odot} \right)^{10/3} \left( \frac{f_{gw}}{100 \text{ Hz}} \right)^{13/3} \left[ \frac{\text{erg}}{\text{s}} \right]. 
 \end{equation}
Now we want to compare the corresponding power for the particle generation verses the Hawking radiation.

\subsection{Crossover frequency $f_\times(M)$}

For this comparison, we determine at which the frequency of gravitational wave the dynamical particle emission equals
the Hawking luminosity by setting
$P_{\mathrm{dyn}}(f_\times,\mathcal{M}) = P_H(M)$. Let us write $P_{dyn}$ from the particle 
production due to the dynamics of spacetime with the Hawking radiation as 
\begin{align}
P_{\mathrm{dyn}} &= A\,
\left(\frac{\mathcal{M}}{M_\odot}\right)^{10/3}
\left(\frac{f}{100\,\mathrm{Hz}}\right)^{16/3},
 A \sim 10^{-30}\,\mathrm{erg\,s^{-1}},
\\[4pt]
P_H &= B\,\left(\frac{M_\odot}{M}\right)^{2},
\hspace{1cm} B \approx 9.0\times 10^{-22}\,\mathrm{erg\,s^{-1}}.
\end{align}
Equating the two and solving for $f_\times$,
\begin{equation}
f_\times(M) \approx 8.1\,\mathrm{kHz}\;(\frac{M}{M_\odot})^{-1}\,.
\end{equation}
Just before the merger, the binary reaches the innermost stable circular orbit (ISCO) with a frequency of $f_{ISCO}=c^3/(\pi 6^{3/2} GM)$. For a system parameterized by this mass scale, the peak gravitational wave frequency is approximately \cite{Maggiore:2007ulw}:
\begin{equation}
    f_{ISCO} \simeq 4.38 \times 10^{19} \text{ Hz} \left( \frac{2 \times 10^{14} \text{ kg}}{M} \right),
\end{equation}
That is interesting to know the ratio of these two frequencies as 
\begin{equation}
\frac{f_\times}{f_{\mathrm{ISCO}}} \approx \frac{8.1}{4.4} \approx 1.8 .
\end{equation}
Because both $f_\times$ and $f_{\mathrm{ISCO}}$ scale as $M^{-1}$, this ratio
is \emph{mass-independent} and that is larger than one, means that the crossover always lies above ISCO 
which never realize. 

So in the weak field approximation of particle production, the Hawking radiation dominates the flux from the 
particle production.

\section{Conclusion and Discussion}
\label{conc}

In this paper, we investigated the non-thermal particle production of a quantized massless scalar field induced by the dynamically evolving background of an inspiraling binary black hole. By treating the binary’s gravitational field as a classical, time-dependent perturbation $h_{\mu\nu}$ within the standard quadrupole approximation, we calculated the resulting particle creation using two independent frameworks: the Bogoliubov transformation method and the first-order S-matrix formalism. We demonstrated that these two approaches yield identical transition amplitudes, establishing the theoretical consistency of the particle production mechanism.

Our results show that the time-varying gravitational field of the binary excites the quantum vacuum, leading to a spectrum governed by the resonance condition $\omega + \omega' = \omega_{gw}(t)$. The resulting emission is distinctly non-thermal and follows specific power-law scalings driven by the binary's chirping frequency, with  the power emission as $\propto \omega_{gw}^{16/3}$. 

While our analytic treatment is restricted to the inspiral regime where linear perturbation theory holds, the power-law dependence clearly indicates that the effect will be most pronounced as the binary approaches merger. This opens up intriguing astrophysical implications; extending this formalism to electromagnetic fields may provide a novel mechanism for explaining transient phenomena \cite{Petroff:2019}, particularly in the context of low-mass binary black hole mergers. Future work will be required to probe the highly non-linear merger regime and explicitly calculate the observable electromagnetic signatures of this quantum effect.

\begin{acknowledgements}
\end{acknowledgements}

\end{document}